\documentclass[journal]{IEEEtran}
\usepackage{amssymb}
\usepackage{amsfonts}
\usepackage{epstopdf}
\usepackage{graphicx}
\usepackage{stmaryrd}
\usepackage{amssymb,amsmath}
\usepackage{multirow,url}
\usepackage{color,cite}
\usepackage{hyperref} 

\ifodd 1
\newcommand{\com}[1]{\textbf{\color{red} (COMMENT: #1)}} 
\newcommand{\comg}[1]{\textbf{\color{green} (COMMENT: #1)}}
\newcommand{\response}[1]{\textbf{\color{magenta} (RESPONSE: #1)}} 
\else

\newcommand{\com}[1]{}
\newcommand{\comg}[1]{}
\newcommand{\response}[1]{}
\fi
\ifodd 0
\newcommand{\referred}[1]{\textcolor{red}{RefPaper: #1}} 
\else
\newcommand{\referred}[1]{}
\fi

\ifodd 1
\newcommand{\changeblue}[1]{\textcolor{blue}{Modified: #1}} 
\else
\newcommand{\changeblue}[1]{}
\fi

\begin{document}

\title{Network-Coded Multiple Access}

\author{Lu~Lu,~\IEEEmembership{Member,~IEEE,}
        Lizhao~You,~\IEEEmembership{Student Member,~IEEE,} \\
        and~Soung~Chang~Liew,~\IEEEmembership{Fellow,~IEEE}
\IEEEcompsocitemizethanks{\IEEEcompsocthanksitem L. Lu is with the Institute of Network Coding, The Chinese University of Hong Kong, Hong Kong Special Administrative Region, China. E-mail: lulu@ie.cuhk.edu.hk
\IEEEcompsocthanksitem L. You and S. C. Liew are with the Department of Information Engineering, The Chinese University of Hong Kong, Hong Kong Special Administrative Region, China. E-mails: \{lzyou, soung\}@ie.cuhk.edu.hk
}
\thanks{}}

\maketitle

\begin{abstract}
This paper proposes and experimentally demonstrates a first wireless local area network (WLAN) system that jointly exploits physical-layer network coding (PNC) and multiuser decoding (MUD) to boost system throughput. We refer to this multiple access mode as Network-Coded Multiple Access (NCMA). Prior studies on PNC mostly focused on relay networks. NCMA is the first realized multiple access scheme that establishes the usefulness of PNC in a \emph{non-relay setting}. NCMA allows multiple nodes to transmit simultaneously to the access point (AP) to boost throughput. In the non-relay setting, when two nodes A and B transmit to the AP simultaneously, the AP aims to obtain both packet A and packet B rather than their network-coded packet. An interesting question is whether network coding, specifically PNC which extracts packet $A \oplus B$, can still be useful in such a setting. We provide an affirmative answer to this question with a novel two-layer decoding approach amenable to real-time implementation. Our USRP prototype indicates that NCMA can boost throughput by 100\% in the medium-high SNR regime ($\ge$10dB). We believe further throughput enhancement is possible by allowing more than two users to transmit together.
\end{abstract}



\section{Introduction}

\IEEEPARstart{T}his paper proposes and experimentally demonstrates a first wireless local area network (WLAN) system that jointly exploits physical-layer network coding (PNC) \referred{PNC06}\cite{PNC06} and multiuser decoding (MUD) \referred{Verdubook}\cite{Verdubook} to boost system throughput. We refer to this multiple access mode as Network-Coded Multiple Access (NCMA).

Since its introduction in \referred{PNC06,popovski2006anti}\cite{PNC06,popovski2006anti}, PNC has developed into a subfield of network coding with a wide following \referred{Nazer2011ReliablePNC, PNCSurveyPhycom12}\cite{Nazer2011ReliablePNC,PNCSurveyPhycom12}. Nearly all prior studies of PNC, however, focused on \emph{relay} networks. NCMA, as expounded in this paper, is the first realized multiple access scheme that establishes the usefulness of PNC in a \emph{non-relay setting}.

\begin{figure}
\centering
\includegraphics[width=0.5\textwidth]{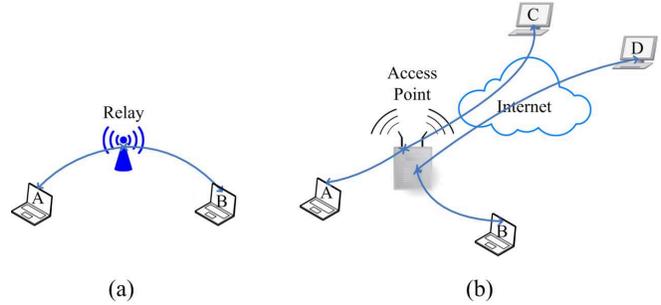}
\caption{Traffic patterns of (a) a two-way relay network; (b) a multiple-access WLAN.}\label{fig:NCMA-Model}
\end{figure}

To put things in context, let us first review the application of PNC in a two-way relay network (TWRN), as shown in Fig. \ref{fig:NCMA-Model}(a).  Here, two nodes A and B wish to send messages to each other via a relay R. With PNC, nodes A and B send their packets, $A$ and $B$, to relay R simultaneously. Relay R then derives a network-coded packet (e.g., a bit-wise XOR packet $A \oplus B$) from the received overlapping signals. It then broadcasts $A \oplus B$ to nodes A and B. With $A \oplus B$, node A recovers packet $B$ using self-information: $B = A \oplus (A \oplus B)$; likewise for node B. In this way, only two time slots are needed for the two-packet exchange, and 100\% throughput improvement can be achieved with respect to the 4-time-slot traditional relaying scheme \referred{PNC06}\cite{PNC06}.


In a WLAN setting, as shown in Fig. \ref{fig:NCMA-Model}(b), instead of a relay, we have an access point (AP). Oftentimes, nodes A and B are not interested in communicating with each other. Rather, they want to communicate with some other nodes, say in the Internet, as shown in Fig. \ref{fig:NCMA-Model}(b). In this case, the communicating counterpart of A is C, not B. For the AP to relay packet $A \oplus B$ to node C is not useful because C lacks packet $B$. For compatibility with the legacy Internet, the AP has to forward the individual messages of A and B to their respective destinations. As far as the communication within the WLAN is concerned, we have a \emph{non-relay setting} in which the AP serves as the termination point for the messages from A and B in that its ultimate goal is to derive these individual messages explicitly rather than their network-coded message. There is no relaying between nodes A and B within the WLAN.

An interesting question is whether packet $A \oplus B$ is still useful in this non-relay setting. We provide an affirmative answer to this question with a novel two-layer decoding approach amenable to real-time implementation. Two key enabling components are our specially designed 1) PHY-layer channel decoders, and 2) MAC-layer erasure channel decoders.

For component 1), we first use a \emph{MUD channel decoder} to try to decode both packets $A$ and $B$. If the MUD decoder successfully decodes both packets, our job is done. However, our experimental data indicate that sometimes only packet $A$ or packet $B$ can be obtained, and sometimes none of them. In the event that the MUD decoder can decode only one of the packet $A$ or $B$, or none of them, we use a \emph{PNC channel decoder} to try to decode $A \oplus B$, The likelihood of the PNC decoder successfully decoding $A \oplus B$ when the MUD decoder does not have complete success in decoding packets $A$ and $B$ can be substantial. For example, at SNR of 8.5dB, with probability 22\% the MUD decoder can decode only one of packet $A$ or $B$. When only one of the packets can be decoded, with probability 85\% we can decode $A \oplus B$. In this scenario, we can use $A \oplus B$ and the available native packet to recover the missing native packet, $A$ or $B$. We refer to this as the \emph{PHY-layer bridging} mechanism of network-coded packets. A native packet can be used with a \emph{complementary network-coded packet} to recover a missing native packet. This is the first way the network-coded packet can be useful.

Component 2) provides a second way network-coded packets can be useful. Our experimental data indicates that at SNR of 8.5dB, with probability 55\% the MUD decoder fails to decode both packets $A$ and $B$. When both $A$ and $B$ cannot be decoded, with probability 40\%, the PNC decoder can still decode $A \oplus B$. At first glance, this lone $A \oplus B$ is not useful because we cannot use it together with an available native packet to recover a missing native packet. An interesting question is whether we can still extract utility out of such \emph{lone network-coded packets}. Seemingly, the answer is no because there is no mutual information between the network-coded packet with either of the two native packets when the other native packet is not available.


Component 2), however, provides a way to use the lone network-coded packets by exploiting mutual information at the \emph{message} level. At the MAC layer of NCMA, block messages $M^A$ and $M^B$ from A and B are coded using an erasure channel code (e.g., the Reed Solomon code \referred{RScodes}\cite{RScodes}) and partitioned into smaller constituent packets. For example, the block messages $M^A$ and $M^B$ could be jumbograms from IPv6 or other large messages from the network layer (e.g., those from big data communication \referred{BigData}\cite{BigData}). With erasure channel coding, provided enough of the constituent packets of $A$ (or $B$) can be decoded at the PHY layer, then $M^A$ (or $M^B$) can be obtained at the MAC layer. Here, we give a simple example on how the lone packets from the PHY layer can be useful. Suppose that at some point in time, the PHY layer has decoded enough packets of A for it to obtain $M^A$. Having the source message $M^A$ then allows the AP to derive all the missing packets $A$ in the previous PHY transmissions. This includes the time slots in which the PHY decoders could not decode both packets $A$ and $B$ but could decode packet $A \oplus B$. With the newly derived packets $A$, their corresponding missing packets $B$ can now be recovered through the previously lone $A \oplus B$. We see that once one of $M^A$ or $M^B$ is decoded, the lone $A \oplus B$ packets become complementary and useful. We refer to this as the \emph{MAC-layer bridging} mechanism of network-coded packets.

The contributions of this work are as follows:
\begin{enumerate}\leftmargin=0in 
\item We present the first conceptualization and experimental demonstration of the usefulness of PNC in a non-relay setting.
\item In terms of concepts and principles, we
    \begin{enumerate}
    \item design a simple MUD decoder and a simple PNC decoder for based on the principle of \emph{reduced constellation} that are amenable to real-time implementation;
    \item devise PHY-layer and MAC-layer bridging algorithms that fully exploit the information contained in the native and network-coded packets decoded by the PHY decoders;
    \item propose an NCMA MAC protocol that can realize the potential throughput gain of our decoding algorithms.
    \end{enumerate}
\item In terms of experimentation, we
    \begin{enumerate}
    \item prototype our NCMA system on the USRP platform to prove its viability;
    \item show that our NCMA system can achieve 100\% throughput gain at the medium-high SNR regime ($\ge$10dB) compared with the user-by-user transmission system;
    \item demonstrate the robustness of our NCMA system under both balanced and unbalanced receive powers from different users.
    \end{enumerate}
\end{enumerate}


\section{Overview} \label{sec:Overview}
This section gives a quick overview of the architecture of the NCMA system and the experimental results that motivates its design.
\begin{figure}
\centering
\includegraphics[width=0.5\textwidth]{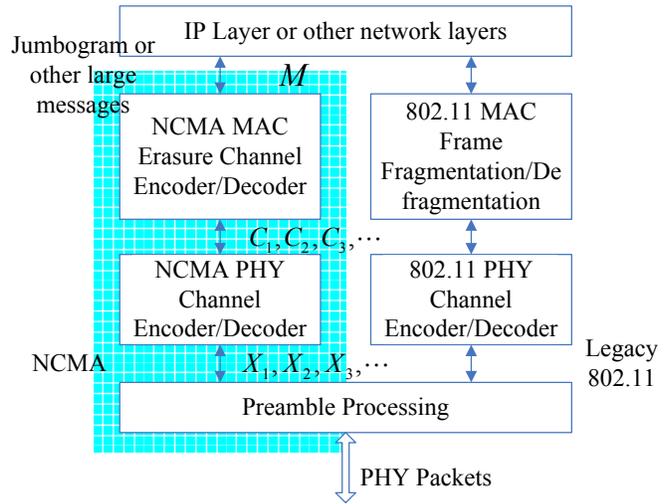}
\caption{The architecture of an NCMA node.}\label{fig:NCMA-Architec}
\end{figure}

\subsection{NCMA Architecture} \label{sec:Overview1}
Fig. \ref{fig:NCMA-Architec} shows the architecture of a node equipped with NCMA capability. For legacy compatibility, the node can revert to 802.11 when necessary. NCMA packets use different preambles than 802.11 so that the receiver can distinguish between NCMA packets and 802.11 packets.

NCMA can support IPv6 jumbograms or other large messages from the network layer. Instead of chopping a large message $M$ from the network layer into independent packets, it makes use of an erasure channel code to encode the message into multiple packets, $C_1, C_2, \ldots$. The erasure channel code adopted could be the Reed Solomon (RS) code \referred{RScodes}\cite{RScodes} or a rateless channel code \referred{MackayFountainCodes}\cite{MackayFountainCodes}. Provided a sufficient number of these packets are received correctly, then the original source message can be decoded at the receiver. At the PHY layer, each packet $C_i$ is further channel-coded into a packet $X_i$. For our USRP prototype, we adopt the RS code at the MAC layer and the convolutional code used in 802.11 at the PHY layer.

In the NCMA mode, the node transmits packets $X_1, \\ X_2,  \dots$ in different time slots and a packet will not be retransmitted even if it cannot be received successfully. Also, there is no acknowledgement from the receiver until it has successfully decoded the associated message $M$.

\subsection{PHY-Layer Decoding and Bridging} \label{sec:Overview2}

\begin{figure}
\centering
\includegraphics[width=0.45\textwidth]{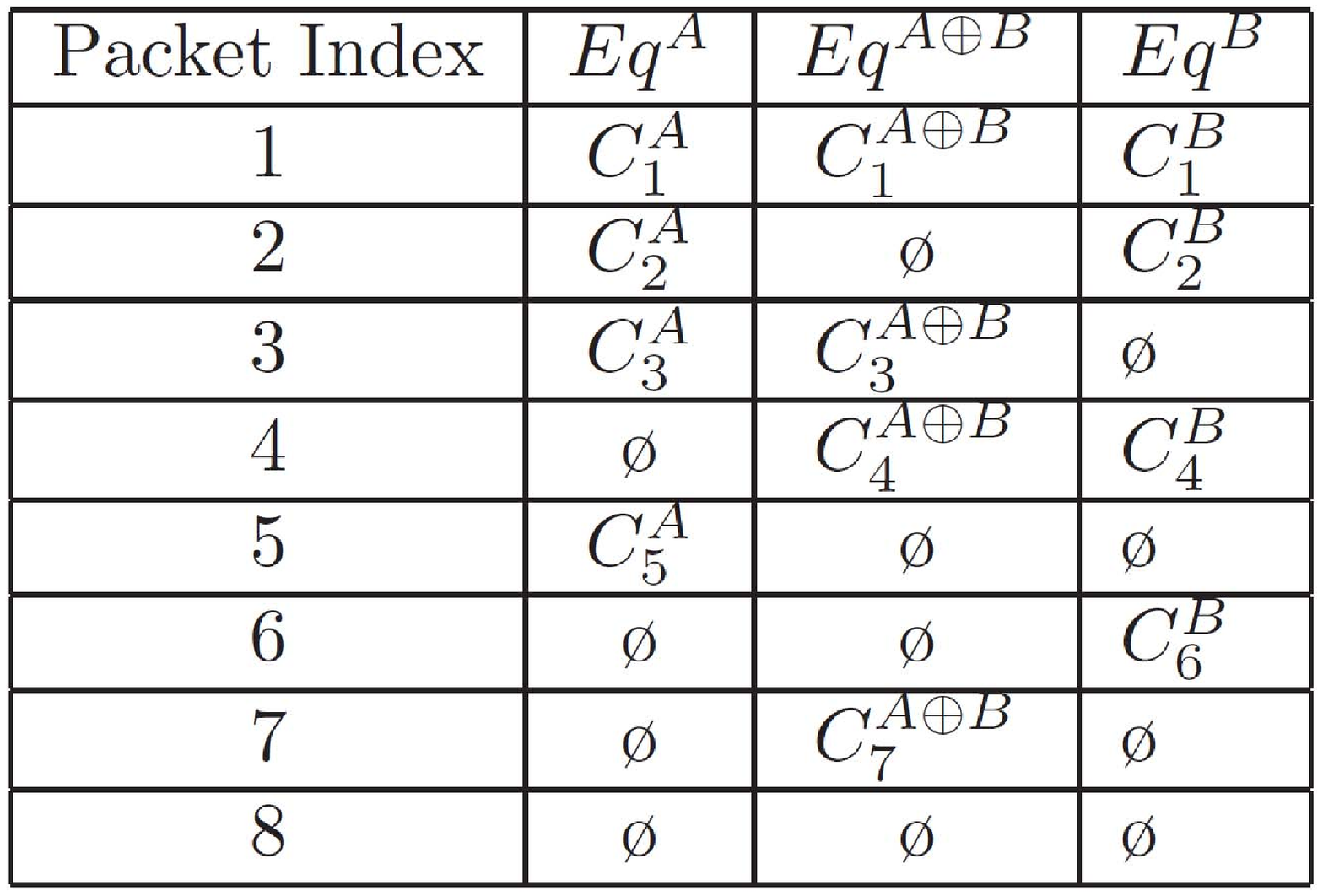}
\caption{An example of PHY-layer packet reception patterns for concurrently transmitted packets. Empty entries (\o) mean the corresponding packets cannot be decoded. Each column is labeled $Eq^J$ corresponding to packet type $J$ being decoded. Each column forms a linear equation system for MAC-layer decoding (see Section \ref{sec:MacDec}).}\label{fig:NCMA-Event}
\end{figure}

We adapt the standard single-user soft input Viterbi decoding algorithm (VA) \referred{forney1973viterbi}\cite{forney1973viterbi} for use in NCMA. Adaptations are necessary because NCMA is a multiuser system rather that a single-user system. The details on how we adapt the standard single-user VA for PNC and MUD decoding can be found in Section \ref{sec:PhyDec}. Importantly, we emphasize that this adaptation approach aims for reduced complexity rather than optimality in the PNC and MUD decoder designs. More optimal decoders are certainly possible; however, the standard VA decoder is a very simple decoder that allows efficient real-time decoding (see Section \ref{sec:PhyDec}).

We make use of two multiuser PHY-layer channel decoders: a MUD decoder and a PNC decoder. When two nodes A and B transmit to the AP simultaneously, the MUD decoder attempts to decode both $C_i^A$ and $C_i^B$ based on the overlapped signals of $X_i^A$ and $X_i^B$; the PNC decoder, on the other hand, attempts to decode $C_i^A  \oplus C_i^B$ (the bit-wise XOR of $C_i^A$ and $C_i^B$) based on the same overlapped signals.


\emph{\textbf{Eight Possible Decoding Outcomes}} -- For the MUD decoder, there are four possible outcomes: (i) both $C_i^A$ and $C_i^B$ are successfully decoded; (ii) only $C_i^A$ is successfully decoded; (iii) only $C_i^B$ is successfully decoded; (iv) both $C_i^A$ and $C_i^B$ cannot be decoded. For the PNC decoder, there are two possible outcomes: (I) $C_i^A \oplus C_i^B$ is successfully decoded; ($\textnormal{II}$) $C_i^A \oplus C_i^B$ cannot be decoded. As a result we have $4 \times 2 = 8$ possible combined outcomes. Fig. \ref{fig:NCMA-Event} shows a contrived example in which the eight possible combined outcomes (events) occur in successive time slots.




\subsubsection{Leveraging Complementary XOR Packets}

In Fig. \ref{fig:NCMA-Event}, event (ii)(I) and event (iii)(I) occur in time slots 3 and 4, in which $C_3^A$ and $C_3^A \oplus C_3^B$ (abbreviated as $C_3^{A \oplus B}$), and $C_4^B $ and $C_4^A \oplus C_4^B$ (abbreviated as $C_4^{A \oplus B}$), are decoded, respectively. In these two cases, the complementary XOR packets, $C_3^A \oplus C_3^B$ and $C_4^A \oplus C_4^B$, can be used to recover the missing native packets, $C_3^B$ and $C_4^A$, respectively. That is, these two events are equivalent to event (i) in which both native packets can be decoded.

Let us take a glipmse of the experimental results of Fig. \ref{fig:NCMA-Exp-PHY-Distribution1}, the details of which will be discussed in Section \ref{sec:Exp}. At SNR=8.5dB, we see that the probabilities of events (ii)(I) and (iii)(I), annotated as \texttt{AX|BX} in the figure, are not negligible. Specifically, they occur around 20\% of the time. This means the complementary XOR packets can help recover the missing native packets around 20\% of the time.




\subsection{MAC-Layer Decoding and Bridging}\label{sec:Overview3}

Returning to Fig. \ref{fig:NCMA-Event}, event (iv)(I) occurs in time slot 7. None of the native packets can be decoded by the MUD decoder; on the other hand, the PNC decoder can decode $C_7^A \oplus C_7^B$. Such lone $C_i^A \oplus C_i^B$ are not useful as far as the recovery of the native packets is concerned. Yet, experimental data indicate that event  (iv)(I) is not negligible. As shown in Fig. \ref{fig:NCMA-Exp-PHY-Distribution1}, the event of lone XOR packet, annotated as \texttt{X} in the figure, occurs around 20\% of the time at SNR=8.5dB. This suggests that the system performance could be improved by much if we can find ways to exploit the lone XOR packets.

\begin{figure}[t]
\centering
\includegraphics[width=0.5\textwidth]{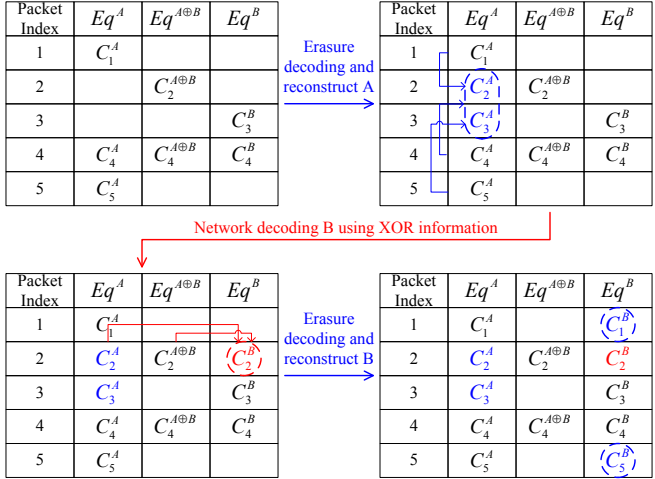}
\caption{NCMA decoding process, with $L= 3$ in the RS code as an example.}\label{fig:NCMA-Overview-MAC-Example}
\end{figure}

Lone $C_i^A  \oplus C_i^B$ turns out to be useful for MAC-layer decoding, where the correlations and mutual information among successive packets can be exploited. Fig. \ref{fig:NCMA-Overview-MAC-Example} illustrates the idea.

\subsubsection{Leveraging Lone XOR Packets}

In the upper left table of Fig. \ref{fig:NCMA-Overview-MAC-Example}, we suppose that the AP has recovered enough PHY packets $C_i^A$ of different $i$ for it to decode $M^A$ -- in this simplified example, $L = 3$ PHY packets are needed. Once $M^A$ is decoded, all the $C_i^A$ that could not be obtained by the PHY-layer decoders can now be recovered (through re-encoding based on $M^A$). In Fig. \ref{fig:NCMA-Overview-MAC-Example}, these are the $C_2^A$ and $C_3^A$ circled in blue as shown in the upper right table. Note that once $C_2^A$ is recovered, the previously lone $C_2^{A} \oplus C_2^{B}$ is converted into a complementary XOR packet and can be used to recover $C_2^B$, shown in red circle in the lower left table. In this example, we now also have enough $C_i^B$ of different $i$ to decode $M^B$. The details and effectiveness of such MAC-layer bridging by XOR packets will be addressed in Section \ref{sec:MacDec} and Section \ref{sec:Exp}, respectively.


\subsection{NCMA MAC Protocol}\label{sec:Overview4}

To exploit the aforementioned decoding and bridging mechanisms, we need a MAC protocol that promotes concurrent transmissions by multiple nodes to the AP. There are many possibilities. Here we give a simple example that covers only the essence of NCMA.

\begin{figure}
\centering
\includegraphics[width=0.5\textwidth]{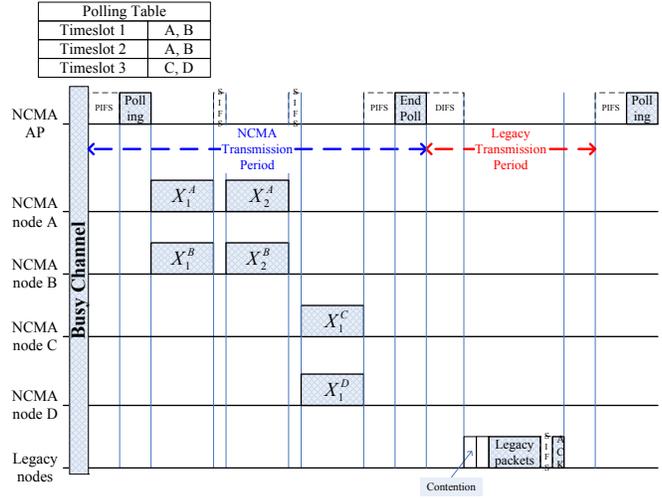}
\caption{NCMA uplink media access with NCMA and legacy nodes. The polling table is contained in the \emph{Polling} frame, and the \emph{End Poll} frame terminates the NCMA channel access session. }\label{fig:NCMA-Protocol}
\end{figure}

Fig. \ref{fig:NCMA-Protocol} illustrates the uplink operation of this protocol. Here, there are four NCMA nodes: A, B, C, and D. Their source messages have been encoded into PHY-layer packets: $X_1^A, X_2^A, \dots$; $X_1^B, X_2^B, \dots$; $X_1^C, X_2^C, \dots$; $X_1^D, X_2^D, \dots$. As far as the scheduling of the transmissions is concerned, the NCMA operation is similar to, but not exactly the same as, the 802.11 point coordination function (PCF)\referred{dot11std09}\cite{dot11std09}. As in 802.11 PCF, a transmission round is divided two periods: contention-free period (CFP) and contention period. NCMA nodes transmit during CFP, and legacy 802.11 nodes transmit during the contention period using 802.11 DCF.

The AP coordinates the NCMA transmissions during CFP by polling. Each poll coordinates the transmissions in a number of successive transmission time slots. A difference compared with the PCF in 802.11 is that a number of nodes may transmit together in a time slot here. In other words, instead of avoiding collisions as in the original PCF, here the AP actually tries to encourage ``collisions''. As illustrated in Fig. \ref{fig:NCMA-Protocol}, nodes A and B transmit together in time slots 1 and 2, and nodes C and D transmit together in time slot 3.

Another difference with 802.11 PCF is that the AP does not issue an ACK immediately upon the reception of each packet. Rather, an ACK will be issued only when the whole MAC-layer source message $M$ of a source has been decoded.

\section{MAC-Layer Message Decoding Algorithm}\label{sec:MacDec}

This section delves into the details of NCMA MAC-layer decoding.

\subsection{Preliminary: Single-User Erasure Channel Coding and Decoding} \label{sec:MacDec1}
We first review the single user case in which only node A transmits. For a focus, we assume the use of the RS erasure channel code \referred{RScodes}\cite{RScodes}. Extensions to other erasure codes, including advanced rateless codes \referred{MackayFountainCodes}\cite{MackayFountainCodes}, are possible.

We adopt the set-up as expounded in \referred{LeeLiewATM97}\cite{LeeLiewATM97}. We organize the symbols of source message $M^A$ of node A in matrix form:
\begin{align}
M^A = \left[
\begin{matrix}
  M_1^A \\
  \vdots\\
  M_i^A\\
  \vdots\\
  M_K^A
\end{matrix}
\right]^T
=
\begin{bmatrix} {a_{1,1} } & {a_{1,2} } &  \cdots  & {a_{1,L} } \\
                \vdots  &  \vdots  &  \ddots  &  \vdots \\
              {a_{i,1} } & {a_{i,2} } &  \cdots  & {a_{i,L} }  \\
                \vdots  &  \vdots  &  \ddots  &  \vdots   \\
               {a_{K,1} } & {a_{K,2} } &  \cdots  & {a_{K,L} }  \\
\end{bmatrix},
\label{equ:single1}
\end{align}
where $a_{i,j}  \in GF(2^s )$. For example, if $s = 8$, each symbol $a_{i,j}$ is a byte and there are altogether $KL$ bytes in the source message. The generator matrix for the RS code is
\begin{align}
G^A =
\begin{bmatrix}
   {G_1 }  \\
    \vdots \\
   {G_i }  \\
    \vdots \\
   {G_N }
\end{bmatrix}
=\begin{bmatrix}
   {g_{1,1} } & {g_{1,2} } &  \cdots  & {g_{1,L} }  \\
    \vdots  &  \vdots  &  \ddots  &  \vdots         \\
   {g_{i,1} } & {g_{i,2} } &  \cdots  & {g_{i,L} }  \\
    \vdots  &  \vdots  &  \ddots  &  \vdots         \\
   {g_{N,1} } & {g_{N,2} } &  \cdots  & {g_{N,L} }
\end{bmatrix},
\label{equ:single2}
\end{align}
where $g_{i,j}  \in GF(2^s )$. There are altogether $N = 2^s - 1$ nonzero elements in $GF(2^s)$. Let us denote the nonzero elements by $\alpha _1 ,\alpha _2 , \ldots ,\alpha _{N}$. We set
\begin{align}
G_i  = \begin{bmatrix}
   1 & {\alpha _{_i } } & {\alpha _i^2 } &  \cdots  & {\alpha _i^{L - 1} }
\end{bmatrix},
{{~~~}}1 \le i \le N. \nonumber
\end{align}

\noindent \textbf{Proposition 1}: Any $L$ of the vectors $G_1 ,G_2 , \dots ,G_{N - 1}$ are linearly independent.

Proof: see \referred{LeeLiewATM97}\cite{LeeLiewATM97}.

For each message $M^A$, we can generate $N$ coded packets, each of $K$ symbols, by
\begin{align}
C^A  =
\begin{bmatrix}
   {C_1^A }  \\    \vdots \\  {C_i^A } \\   \vdots \\   {C_N^A }
\end{bmatrix}
= G(M^A)^T.
\label{equ:single3}
\end{align}
Upon receiving any $L$ packets, denoted by $C_{(1)}^A ,C_{(2)}^A, \dots , C_{(L)^A}$, we have:
\begin{align}
\tilde C^A  =
\begin{bmatrix}
   {C_{(1)}^A }  \\
    \vdots       \\
   {C_{(i)}^A }  \\
    \vdots       \\
   {C_{(L)}^A }
\end{bmatrix}
=\tilde G(M^A )^T=
\begin{bmatrix}
   {G_{(1)}^{} }  \\
    \vdots        \\
   {G_{(i)}^{} }  \\
    \vdots        \\
   {C_{(L)}^{} }
\end{bmatrix}
\begin{bmatrix}
   {M_1^A }  \\
    \vdots   \\
   {M_i^A }  \\
    \vdots   \\
   {M_K^A }
\end{bmatrix}^{T} .
\label{equ:single3}
\end{align} 
By Proposition 1, $\tilde G$ is invertible. Thus, we can extract message A by
\begin{align}
(M^A )^T  = \tilde G^{ - 1} \tilde C^A      .
\label{equ:single4}
\end{align}

With $s = 8$, we have $N=255$ coded packets. Recall that in NCMA, we do not have PHY-layer ACK and the same packet $C^A_i$ will not be retransmitted if it cannot be received at the AP. There is a chance that the transmitter runs of packets $C^A_i$ to transmit after all 255 of them have been transmitted and still the AP cannot decode $M^A$. However, in the experiments to be detailed in Section \ref{sec:Exp}, the typical $L$ we need is no more than 32, and the transmitter never runs out of packets to transmit before the message is decoded.

\subsection{NCMA Erasure Channel Coding and Decoding} \label{sec:MacDec2}
We now consider the NCMA situation in which node A and node B transmit their $i$-th packets, $C_i^A$ and $C_i^B$, simultaneously for $i = 1,2, \dots$ until at least one of the two messages, $M^A$ or $M^B$, is decoded. For optimal decoding, we set up three interacting equation systems: the first, denoted by $Eq^A$, is for decoding $M^A$; the second, denoted by $Eq^B$, is for decoding $M^B$; and the third, denoted by $Eq^{A \oplus B}$, is for decoding $M^A  \oplus M^B$.

\subsubsection{Equation System $Eq^A$}
Recall that there are eight possible outcomes (events) (see Section \ref{sec:Overview2} and Fig. \ref{fig:NCMA-Event}) for PHY-layer decoding, corresponding to the decoding outcomes of the MUD decoder and PNC decoder for transmission $i$:

We note that for events, (i), (ii), and (iii)(I), $C_i^A$ can be obtained. For events (i) and (ii) (whether the PNC decoding is successful or not), $C_i^A$ is directly given by the MUD decoder. For event (iii)(I), $C_i^A$ is obtained by $C_i^A  = C_i^B  \oplus C_i^{A \oplus B}$.

Any of the events (i), (ii), or (iii)(I), yields a packet of type $C_i^A$. As nodes A and B transmit their successive packets, as soon as we have $L$ packet of type $C_i^A$, we can decode message $M^A$.

\subsubsection{Equation System $Eq^B$}
Similar to $Eq^A$ above. As soon as we have $L$ packets of type $C_i^B$, we can decode $M^B$.

\subsubsection{Equation System $Eq^{A \oplus B}$}
Here, we are interested in event (i) and event (I), since in both these events $C_i^{A \oplus B}$ can be obtained. As for $Eq^A$, as soon as we have $L$ packets of type $C_i^{A \oplus B}$, we have enough equations to decode $M^A  \oplus M^B$ using the same strategy. In particular, thanks to the linearity of the RS code, the decoding is similar to (\ref{equ:single4}).

\subsubsection{Interactions of $Eq^A$, $Eq^B$, and $Eq^{A \oplus B}$}

In Section \ref{sec:Overview3}, we quickly went through the example in Fig. \ref{fig:NCMA-Overview-MAC-Example}. Here, we will examine this example from the angle of the interactions between the three equation systems. At the end of time slot 5, the situation is depicted in the upper left table in Fig. \ref{fig:NCMA-Overview-MAC-Example}. The status of the three equation systems is
\begin{align}
Eq^A  &: {~~~~~~~~}
\tilde C^A =
\begin{bmatrix}
   {C_1^A }  \\
   {C_4^A }  \\
   {C_5^A }
\end{bmatrix}
=
\begin{bmatrix}
   {G_1^{} }  \\
   {G_4^{} }  \\
   {G_5^{} }
\end{bmatrix}(M^A )^T \nonumber \\
Eq^B  &: {~~~~~~~~}
\tilde C^B =
\begin{bmatrix}
   {C_3^B }  \\
   {C_4^B }
\end{bmatrix}
=
\begin{bmatrix}
   {G_3^{} }  \\
   {G_4^{} }
\end{bmatrix}(M^B )^T \nonumber\\
Eq^{A \oplus B}  &: {~~~~~~~~}
\tilde C^{A \oplus B} =
\begin{bmatrix}
   {C_2^{A \oplus B} }  \\
   {C_4^{A \oplus B} }
\end{bmatrix}
=
\begin{bmatrix}
   {G_2^{} }  \\
   {G_4^{} }
\end{bmatrix}(M^A  \oplus M^B )^T \nonumber
\end{align}

In this example, $L = 3$. According to Proposition 1, $\left[ \begin{array} {lcr} {G_1 } & {G_4^{} } & {G_5^{}} \end{array} \right]$ is a full-rank matrix. Therefore, we can invert it to solve for $M^A$ in $Eq^A$. Once, we get $M^A$, we can solve for the two missing packets: $C_2^A  = G_2 (M^A )^T $; $C_3^A  = G_3 (M^A )^T $. Exploiting this, the new status is shown in the upper right table in Fig. \ref{fig:NCMA-Overview-MAC-Example}.

We further note that with $C_2^A $ and $C_2^{A \oplus B}$ , we can then get $C_2^B  = C_2^A  \oplus C_2^{A \oplus B} $, with the new status shown in the lower left table. Now, we also have $L = 3$ equations for $Eq^B$, from which we can get $M^B$.

The XOR equations $Eq^{A \oplus B}$ serve as a bridge between $Eq^A$ and $Eq^B$ at the MAC layer. The bridge can increase the number of available equations in $Eq^A$ and $Eq^B$. When one of the two equation systems, $Eq^A$ or $Eq^B$, has $L$ equations, then additional equations in the other equation system can also be obtained through the bridge.


It is also possible that $Eq^{A \oplus B}$ has $L$ equations before $Eq^A$ or $Eq^B$ does. In this case, the availability of $M^A  \oplus M^B$ gives all the packets $C_i^{A \oplus B}$, $i = 1,\dots,N$, which can then serve as the bridge to obtain $C_i^A$ and $C_i^B$ that were previously unavailable. In particular, once $Eq^{A \oplus B}$ is solved, the number of available equations in $Eq^{A}$ and $Eq^{B}$ will be the same from then on, because knowing $C_i^{A}$ means knowing $C_i^{B}$, and vice versa. Thus, once $Eq^{A \oplus B}$ is solved, either both $Eq^{A}$ and $Eq^{B}$ are solved at the same time as $Eq^{A \oplus B}$ (when the bridge causes them both to have $L$ or more equations), or both $Eq^{A}$ and $Eq^{B}$ need the same number of additional equations for solution (when they still have fewer than $L$ equations). For the latter, the subsequent transmissions will be more efficient because each time when only $C_i^A$ or $C_i^B$ is decoded at the PHY layer, the missing packet can be derived through the bridge.


\subsubsection{Intermixing Message Pairs} \label{sec:MacDec25}
In the example of Fig. \ref{fig:NCMA-Overview-MAC-Example}, enough equations are created through the bridge that message $M^B$ can also be solved at the same time as message $M^A$. In general, it is possible that $Eq^B$ still may not have enough equations. Then $M^B$ of node B can be paired with the next message of node A, or be paired with a message from a different node, say node C. This seamless pairing will ensure that the available equations in $Eq^B$ are not wasted. The new message, of course, will start off with needing $L$ equations while $M^B$ needs fewer equations for eventual solution. Note that the packets associated with the new message, say $M^C$, will be sent in the sequence of $C^C_{j+1}, C^C_{j+2}, \dots , C^C_1, C^C_2, \dots$, if $C^B_j$ was the last packet sent by node B. This is so as to align the equations of node B and node C in the three equation systems.


\section{PHY-Layer Channel Decoding}\label{sec:PhyDec}


The preceding section focused on MAC-layer decoding. At the PHY layer, channel coding is applied to $C_i^A$ and $C_i^B$ by node A and node B to form $X_i^A$ and $X_i^B$ for transmission. At the AP, the PHY decoders attempt to obtain $C_i^A$, $C_i^B$, and $C_i^{A \oplus B}$ from the overlapped $X_i^A$ and $X_i^B$.


At the PHY layer, our NCMA prototype adopts the same convolutional code as in 802.11. A goal of ours is to simplify the PHY decoder design so that real-time decoding is possible in our USRP prototype. The standard VA decoder is widely used in real communication system thanks to its low complexity. As such, we adapt the standard soft-input Viterbi decoding algorithm (VA) \referred{forney1973viterbi}\cite{forney1973viterbi} for PNC and MUD decoding. More optimal decoders with better decoding performance are certainly possible for PNC and MUD; however, they are more complex and may not amenable to real-time decoding. An issue, however, is that the standard VA is meant for a single user system. NCMA, by contrast, is intrinsically a multiuser system, and the PNC and MUD decoders have to deal with overlapped signals from multiple users. VA cannot be used as is. This section explains how we adapt the VA for PNC and MUD decoding.


\subsection{Preliminary: Single-User VA} \label{sec:PhyDec1}

We first review the key idea of the standard VA when applied to the single-user system. Without loss of generality, let us assume an OFDM system (our prototype is an OFDM system). Let $X$ denote the the packet transmitted and $x[k]$ denote the value of bit $k$ within the packet $X$. Let $y[k]$ be the received signal corresponding to bit $k$ given by
\begin{align}
y[k] = h_{s_k} x[k] + n[k], \label{equ:systemptop}
\end{align}
where $x[k] = 1$ if bit $k$ is 0 and $x[k] = -1$ if bit $k$ is 1; and $n[k]$ is the zero mean Gaussian noise with variance $\sigma^2$. In the above, note that instead of writing $h$ for the channel gain, we write $h_{s_k}$, where $s_k$ is the subcarrier on which bit $k$ is transmitted, and $h_{s_k}$ is the channel gain associated with that subcarrier.

The idea of the standard VA is to provide a confidence metric for each bit, $\tilde{x}[k]$, to the Viterbi shortest-path algorithm. This confidence metric is also referred to as the soft bit, and is computed by the log likelihood ratio $\log{P_0[k] \over P_1[k]}$, where $P_0[k] = P(x[k]=1 | y[k])$ and $P_1[k]=P(x[k]=-1|y[k])$. Thus, $\tilde{x}[k]$ takes on a real value. The more positive it is, the more confident we are that bit $k$ is 0; and the more negative it is, the more confident we are that bit $k$ is 1. The VA then finds the maximum likelihood (ML) codeword $X = \left\{x[k]\right\}_{k=1,2,\dots}$ with the minimum $- \sum\limits_k {\tilde{x}[k]} x[k]$ (or maximum $\sum\limits_k {\tilde{x}[k]} x[k]$) \referred{SklarDCbook}\cite{SklarDCbook}. Essentially, a shortest-path algorithm is applied on the trellis associated with the convolutional code \referred{SklarDCbook}\cite{SklarDCbook}.




\subsection{Soft Decoders for NCMA} \label{sec:PhyDec2}

The block diagram of the overall NCMA channel decoder is shown in Fig. \ref{fig:NCMA-ChannelDecoder}. Given the input $\left\{y[k]\right\}_{k=1,2,\dots}$, the AP first detects the presence of signals from the users (using correlation on the preambles). If only one user transmits and only the signal from that user is detected, then the standard single-user VA is used (the lowest branch in the block diagram). If there are overlapping signals from two users, then both the MUD decoder and the PNC decoder are used (the upper two branches in the block diagram).

All the decoders use the standard binary Viterbi algorithm for decoding. Our single-user decoder, PNC decoder, and MUD decoder only differ in the ways their demodulators compute the soft information to be fed to the binary Viterbi decoder. In the next two subsections, we will describe how the PNC demodulator and the MUD demodulator compute their soft information.




\subsubsection{PNC Decoder} \label{sec:PhyDec21}

\begin{figure}
\centering
\includegraphics[width=0.5\textwidth]{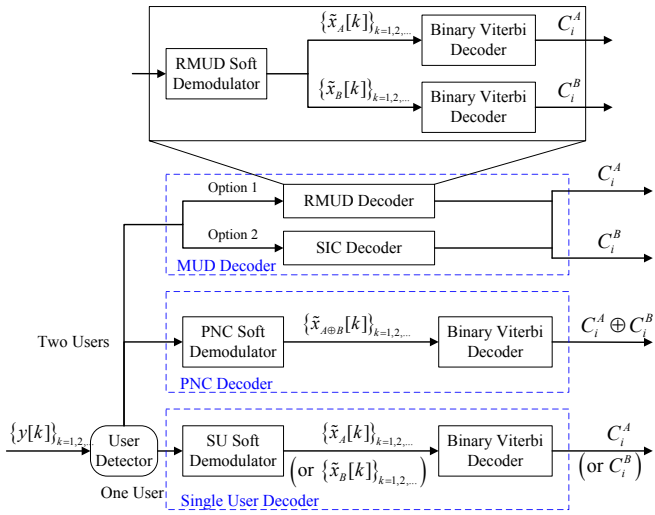}
\caption{Single-user decoder, MUD decoder and PNC decoder designs for overall PHY-layer channel decoding in NCMA. In this figure, the index $k$ is over multiple OFDM symbols in a frame because channel coding is performed over multiple OFDM symbols.}\label{fig:NCMA-ChannelDecoder}
\end{figure}

The main essence of our PNC decoder is as follows. To allow the usage of the standard binary VA, we perform simplification by reducing the number of constellation points to two. These two constellation points correspond to the most probable points for the two different XOR values. The log likelihood ratio based on these two constellation points is used as the soft-information to be fed to the standard binary VA. In Fig. \ref{fig:NCMA-ChannelDecoder}, the PNC demodulator is responsible for computing this soft information. We refer to this strategy as \emph{reduced-constellation decoding}. In the following paragraphs, we describe the details.

Let $X^A = (x_A [1], \dots ,x_A [k], \dots )$ and $X^B = (x_B [1],  \dots, x_B [k], \dots )$ denote the PHY-layer codewords (transmitted packets) of users A and B, and let $\Pi ( \cdot )$ denote functional mapping corresponding to convolutional coding. Since $\Pi ( \cdot )$ is linear, we have $X^A  \oplus X^B  = \Pi (C^A ) \oplus \Pi (C^B ) = \Pi (C^A  \oplus C^B )$. This means that we can first detect the XOR of individual bits $\left\{ {x_A [k] \oplus x_B [k]} \right\}_{k = 1,2,\dots}$ and then feed the information on these XOR bits to the standard convolutional decoder for decoding \referred{PNCSurveyPhycom12}\cite{PNCSurveyPhycom12}\footnote{Note: $x_A[k]$ and $x_B[k]$ adopts value of 1 and $-1$ rather than 0 and 1; therefore, we define $x_A [k] \oplus x_B [k] = x_A [k]x_B [k] $.}. For our PNC decoder, the PNC demodulator will feed soft information on $\left\{ {x_A [k] \oplus x_B [k]} \right\}_{k = 1,2, \dots}$ to the standard VA. For the subsequent discussion, for brevity, we will drop the index $k$ in our notations. In addition, we will write $x_A[k] \oplus x_B[k]$ simply as $x_{A \oplus B}$.


Our proposed $\tilde{x}_{A \oplus B}$, the soft bit for $x_{A \oplus B}$, follows the same reasoning as for $\tilde{x}$ of the standard single-user VA. Recall from the discussion in Section \ref{sec:PhyDec1} that in the standard VA, $\tilde{x} = \log {{P_0} \over {P_1 }}$ is the log likelihood ratio, where $P_0$ is the probability that the bit is 0 and $P_1$ is the probability that the bit is 1.

When there are signals from both nodes A and B, in place of (\ref{equ:systemptop}), the received signal is
\begin{align}
y = h_A x_A + h_B x_B + n,
\label{equ:systempnc}
\end{align}
where $h_A$ and $h_B$ are channel gains of the received packets from nodes A and B respectively, and $n$ is Gaussian noise with variance $\sigma^2$. For $\tilde{x}_{A \oplus B}$, the log likelihood ratio is given by
\begin{align}
\begin{footnotesize}
\begin{array}{l}
\hspace{-0.1in}
  \log {{P_0 } \over {P_1 }} = \log {{\Pr \left\{ {x_A  = 1,x_B  = 1|y} \right\} + \Pr \left\{ {x_A  =  - 1,x_B  =  - 1|y} \right\}} \over {\Pr \left\{ {x_A  =  - 1,x_B  = 1|y} \right\} + \Pr \left\{ {x_A  = 1,x_B  =  - 1|y} \right\}}}  \cr
  {~} = \log \left( {\exp \left\{ { - {{\left| {y - h_A^{} - h_B^{} } \right|^2 } \over {2\sigma ^2 }}} \right\} + \exp \left\{ { - {{\left| {y + h_A^{}  + h_B^{} } \right|^2 } \over {2\sigma ^2 }}} \right\}} \right)  \cr
  {~~~~~} - \log \left( {\exp \left\{ { - {{\left| {y + h_A^{}  - h_B^{} } \right|^2 } \over {2\sigma ^2 }}} \right\} + \exp \left\{ { - {{\left| {y - h_A^{}  + h_B^{} } \right|^2 } \over {2\sigma ^2 }}} \right\}} \right) .
 \end{array}
\label{equ:pncsoft1}
\end{footnotesize}
\end{align}

Unlike the single-user case, it is difficult to further simplify the log likelihood ratio expression above without appromixation. Adopting the log-max approximation, $\log (\sum\nolimits_i {\exp (z_i )} ) \approx \mathop {\max }\limits_i z_i $), \referred{BoydCOBook}\cite{BoydCOBook} gives
\begin{align}
\begin{footnotesize}
\begin{array}{l}
\hspace{-0.2in}
  {{\sigma ^2 } \over 2}\log {{P_0 } \over {P_1 }} \approx {1 \over 4}\max \left\{ { - \left| {y - h_A^{}  - h_B^{} } \right|^2 , - \left| {y + h_A^{}  + h_B^{} } \right|^2 } \right\}  \\
   {~~~~~~~~~~~~} - {1 \over 4}\max \left\{ { - \left| {y + h_A^{}  - h_B^{} } \right|^2 , - \left| {y - h_A^{}  + hs_B^{} } \right|^2 } \right\}.
 \end{array}
\label{equ:pncsoft2}
\end{footnotesize}
\end{align}
We can let $\tilde{x}_{A \oplus B }  = {{\sigma ^2 } \over 2}\log {{P_0 } \over {P_1 }}$ given in (\ref{equ:pncsoft2}) since the constant factor ${{\sigma ^2 } \over 2}$ will not affect the shortest path found by VA.

\begin{figure}
\centering
\includegraphics[width=0.5\textwidth]{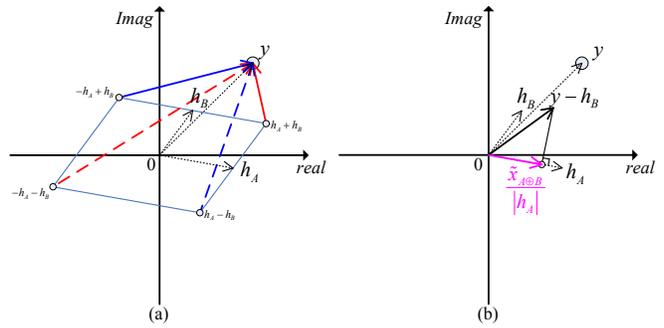}
\caption{Illustration of PNC soft demodulation which gives soft information on XOR bit: the solid and dashed lines of the same color represent the two Euclidean distances between the received sample and the two constellation points mapped to the same XOR. (a) the solid blue line and the solid red line are selected to represent the Euclidean distances to the two different values of the XOR bit $x_{A \oplus B} $; (b) projection of $y- h_B$ onto $h_A$ to get the soft information $\tilde{x}_{A  \oplus B}$.}
\label{fig:NCMA-Soft-PNC-Demod}
\end{figure}

Solution (\ref{equ:pncsoft2}) can be further simplified. We illustrate this with the example in Fig. \ref{fig:NCMA-Soft-PNC-Demod}. Fig. \ref{fig:NCMA-Soft-PNC-Demod} shows the constellation map of a specific pair of channel gains, $h_A$ and $h_B$. With BPSK modulation, $\left( {x_A, x_B } \right)$ takes on four possible values, $( \pm 1, \pm 1)$. In (\ref{equ:pncsoft2}), either $( + 1, + 1)$ or $( - 1, - 1)$ will be chosen in the first $\max(\cdot)$ function, and either $( + 1, - 1)$ or $( - 1, + 1)$ will be chosen in the second $\max(\cdot)$ function. Which pair is chosen depends on $y$. In Fig. \ref{fig:NCMA-Soft-PNC-Demod}, we show a particular realization of $y$ such that $( + 1, + 1)$ and $( - 1, + 1)$ are chosen. The corresponding soft bit for $x_{A \oplus B}$ is
\begin{align}
\hspace{-0.05in}
   \tilde{x}_{A \oplus B}  &\approx  - {1 \over 4}\left| {y - h_A  - h_B } \right|^2  + {1 \over 4}\left| {y + h_A - h_B } \right|^2   \nonumber\\
    &= h_A^{} \cdot \left( {y - h_B^{} } \right).
\label{equ:pncsoft3}
\end{align}

The intuition for (\ref{equ:pncsoft3}) is as follows. The fact that $( + 1, + 1)$ and $( - 1, + 1)$ are chosen implicitly means that we have already made a first decision that node B transmits 1. Thus, we should then decide the XOR value based on what node A transmits. The term $y - h_B$ in (\ref{equ:pncsoft3}) corresponds to subtracting from the received sample $y$ the decision that node B transmits 1. The component that contains the signal from node A can therefore be defined as $y_A  = y - h_B$. It can be shown that the dot product $h_A \cdot y_A $ is simply the soft information for an equivalent single user case in which only node A transmits\footnote{Note that for the single user case, when BPSK is used, the component of $y_A$ that is orthogonal to $h_A$ contains noise only. The dot product in the complex plane extracts the component of $y_A$ that contains signal from node A.}.

For other realizations of $y$, $\tilde{x}_{A \oplus B}$ can be found similarly. Specifically, there are four cases. If $y$ is such that:
\begin{itemize}\leftmargin=0in
\item $( + 1, + 1)$ and $( - 1, + 1)$  are chosen, then $\tilde{x}_{A \oplus B }  \approx  h_A \cdot \left( {y - h_B} \right)$;
\item $( - 1, - 1)$ and $( - 1, + 1)$  are chosen, then $\tilde{x}_{A \oplus B }  \approx -h_B \cdot \left( {y + h_A} \right)$;
\item $( + 1, + 1)$ and $( + 1, - 1)$  are chosen, then $\tilde{x}_{A \oplus B }  \approx  h_B \cdot \left( {y - h_A} \right)$;
\item $( - 1, - 1)$ and $( + 1, - 1)$  are chosen, then $\tilde{x}_{A \oplus B }  \approx -h_A \cdot \left( {y + h_B} \right)$.
\end{itemize}

Feeding $\left\{\tilde{x}_{A \oplus B}[k]\right\}_{k=1,2,\dots}$ to the standard VA allows us to decode for $C^A \oplus C^B$.

\subsubsection{MUD Decoders} \label{sec:PhyDec22}

In this paper, we consider two possible MUD decoders, as shown in the upper portion of Fig. \ref{fig:NCMA-ChannelDecoder}. The first MUD decoder is similar to our PNC decoder in that it is based on the principle of reduced constellation. We refer to this MUD decoder as the reduced-constellation MUD (RMUD) decoder. The second MUD decoder is based on the well-known successive interference cancellation (SIC) principle \referred{SICMobicom07}\cite{SICMobicom07}. We detail these two MUD decoders in the following.

\begin{figure}
\centering
\includegraphics[width=0.5\textwidth]{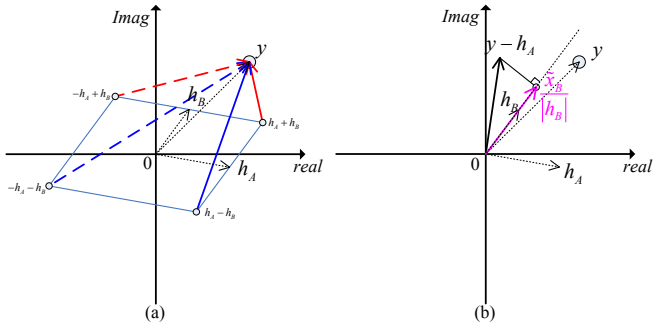}
\caption{Illustration of reduced-constellation MUD soft demodulation for $x_B$: (a) the solid blue line and the solid red lines are selected to represent the Euclidean distances to the two different values of the bit $x_B$; (b) projection of $y - h_A$ onto $h_B$ to get the soft information $\tilde{x}_B$.}
\label{fig:NCMA-Soft-MUD-Demod}
\end{figure}

\vspace{0.05in}
\noindent\textbf{RMUD:}
The RMUD soft demodulator (see Fig. \ref{fig:NCMA-ChannelDecoder}) provides the soft information on $x_A$ and $x_B$, $\tilde{x}_A$ and $\tilde{x}_B$, to two separate standard VA decoders, one for decoding $C_i^A$ and one for decoding $C_i^B$. Without loss of generality, let us focus on $\tilde{x}_B$. The associated log likelihood function is
\begin{align}
\begin{footnotesize}
\begin{array}{l}
\hspace{-0.1in}
 \log {{P_0 } \over {P_1 }}   = \log \left( {\exp \left\{ { - {{\left| {y - h_A^{}  - h_B^{} } \right|^2 } \over {2\sigma ^2 }}} \right\} + \exp \left\{ { - {{\left| {y + h_A^{}  - h_B^{} } \right|^2 } \over {2\sigma ^2 }}} \right\}} \right)  \\
  {~~~~~~~~~} - \log \left( {\exp \left\{ { - {{\left| {y - h_A^{}  + h_B^{} } \right|^2 } \over {2\sigma ^2 }}} \right\} + \exp \left\{ { - {{\left| {y + h_A^{}  + h_B^{} } \right|^2 } \over {2\sigma ^2 }}} \right\}} \right).
 \end{array}
\label{equ:mudsoft1}
\end{footnotesize}
\end{align}
For the same example as in Fig. \ref{fig:NCMA-Soft-PNC-Demod}, (\ref{equ:mudsoft1}) can be simplified to the following using the same strategy as in the preceding section:
\begin{align}
\tilde{x}_B   \approx h_B \cdot \left( {y - h_A} \right).
\label{equ:mudsoft2}
\end{align}
The procedure for getting $\tilde{x}_B$ is illustrated in Fig. \ref{fig:NCMA-Soft-MUD-Demod}.



Similarly, we can get $\tilde{x}_A  \approx h_A \cdot \left( { y - h_B} \right)$, whose expression happens to be the same as that of $\tilde{x}_{A  \oplus B} $ in (\ref{equ:pncsoft3}) for this particular example, but with different physical meanings. As for the PNC modulator in the preceding section, in general there are also four cases to be considered for each of $\tilde{x}_A$ and $\tilde{x}_B$ (omitted here to conserve space).



\vspace{0.05in}
\noindent\textbf{SIC:}
With respect to (\ref{equ:systempnc}), SIC tries to first decode a stronger signal, say $x_A$, and then substitutes the estimate for $x_A$, $\hat{x}_A$, into (\ref{equ:systempnc}) to get $\tilde{y} = {\left(y - h_A \hat{x}_A \right)} / {h_B} = x_B + {h_A \left(x_A - \hat{x}_A\right)}/ {h_B} + \left(n/h_B\right)$. If the overall codeword $\hat{X}^A$ is a valid codeword so that $\hat{X}^A = X^A$, then $y= x_B + \left(n/h_B\right)$. Thus the decoding of $X^B$ would be as if $X^A$ did not exist. If, on the other hand, $X^A$ is not decoded correctly, then $X^B$ may likely suffer decoding errors as well.



Although in principle SIC should decode the stronger signal first, we find that in practice, based on our experimental data, better performance can be obtained by running two parallel SICs, with one decoding signal $A$ first and the other decoding signal $B$ first. It is perhaps due to other distortions and imperfection in channel estimation (in addition to noise) that decoding the stronger signal first is not always the best strategy (although most of time it is). The SIC experimental results that will be presented in Section \ref{sec:Exp} are based on this parallel version.

\subsection{Quantization of Soft Information}

So far we have assumed the soft bit, $\tilde{x}$, is a real number. In actual implementation, $\tilde{x}$ needs to be quantized before using VA. In particular, the VA decoder adopted in our system is based on the \emph{Spiral} Viterbi software generator \referred{SpiralSOVA}\cite{SpiralSOVA} that only accepts 8-bit inputs (from 0 to 255).

In mapping a real $\tilde{x}$ to an 8-bit quantized $\tilde{x}_{quantized}$, the main issue is where to put the constellation points of the signal within the quantized interval [0, 255], and generally there is a trade-off between clipping and quantization errors. We employ a design that optimizes the trade-off point for our multiuser system. Appendix \ref{sec:appendix1} provides the details on such quantization.

\section{Implementation and Experimental Results} \label{sec:Exp}

This section presents the details of our NCMA implementation over the USRP software radio platform \referred{EttusUSRP}\cite{Ettus} and the experimental results.

\subsection{Implementation}\label{sec:Exp1}

We implement the full NCMA PHY-layer decoding algorithms. Both the MUD channel decoder and the PNC channel decoder (see Fig. \ref{fig:NCMA-ChannelDecoder}) with real-time performance have been implemented. To achieve real-time decoding, we have simplified the decoder designs with approximations as explained in Section \ref{sec:PhyDec}.

Our system makes use of USRP hardware \referred{EttusUSRP}\cite{Ettus} and GNU Radio software \referred{GNURadio}\cite{GNURadio} with the UHD hardware driver. We extended the RawOFDM single-user point-to-point OFDM transceiver software \referred{RawOFDMcode}\cite{RawOFDMcode} for the NCMA system. The extensions include:
\begin{itemize}\leftmargin=0in
\item [a)] Modifications of the single-user VA software for PHY-layer channel decoding in NCMA.
\item [b)] Modifications of the preamble and pilot designs. Different user nodes use orthogonal preambles and frequency-domain pilots so as to enable multiuser signal presence detection, multiuser channel estimation, and multiuser CFO tracking and compensation at the AP.
\item [c)] Partial precoding. Partial precoding at the user transmitters has been implemented to reduce the relative CFO between the two users at the AP. The transmitters of the two users make use of the preambles in the poll frames from the AP to estimate CFOs for precoding purposes. The relative CFO is small as a result ($\sim$100 -- 200Hz in our experiments).
\item [d)] CRC checking for PNC systems. The mathematics of CRC check for the single-user system with the 802.11 CRC design \referred{dot11std09}\cite{dot11std09} cannot be directly applied to the PNC case, and we modified the CRC for PNC error detection.
\item [e)] Polling mechanism. The AP uses beacons to poll users to transmit either singly or simultaneously. The signals from multiple users can reach the AP with arrival-time offset that is within the cyclic prefix (CP) of OFDM. Doing so can eliminate OFDM symbol offset between users, thus simplify design \referred{FPNCPhycom12, FICA10, SourceSync2010}\cite{FPNCPhycom12, FICA10, SourceSync2010}. Time synchronization to within $10^{-6}$s has been achieved.
\end{itemize}

Due to page limit, we omit the details of these modifications in this paper except for a), which has already been explained in Section \ref{sec:PhyDec}.

\begin{figure}
\centering
\includegraphics[width=0.5\textwidth]{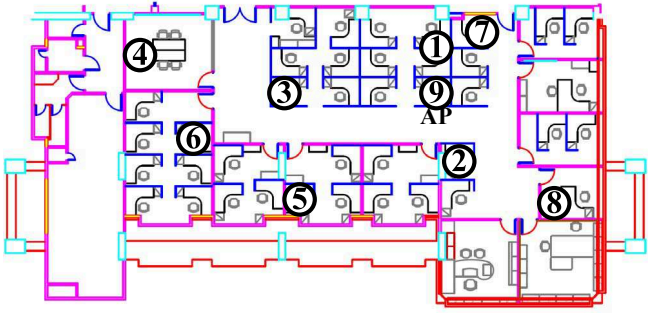}
\caption{Layout of an indoor environment for the deployment of 9 USRP N210 nodes for experiments.}
\label{fig:NCMA-Exp-Map}
\end{figure}

\subsection{Experimental Setup and Results}\label{sec:Exp2}

For experimentation, we deployed 9 sets of USRP N210 with XCVR2450 boards \referred{EttusUSRP}\cite{Ettus} indoor to emulate a WLAN system. The topology is shown in Fig. \ref{fig:NCMA-Exp-Map}. Each node is a USRP-connected PC. Each of the 9 nodes can be chosen to serve as the AP to test different network configurations. The AP can poll any two of the remaining 8 nodes to transmit together. BPSK modulation is used. Our experiments were carried out at 802.11 channel 1 (i.e., 2.412GHz) with 4 MHz bandwidth at midnight to minimize the co-channel interference from nearby ISM band equipment.


To benchmark our NCMA system, we consider the following systems:
\begin{enumerate}\leftmargin=0in
\item \emph{Single-User (SU) system} \\
    This is the traditional user-by-user non-overlapped transmission system. We use the same beacon mechanism as in NCMA to poll a pair of users; however, one user delays its transmission until the other user finishes transmission.
\item \emph{MUD system (Multi-User)} \\
    Here, we only use the MUD decoder (and not the PNC decoder) for PHY-layer decoding. The MUD decoder can either be RMUD or SIC. Only two equation systems, $Eq^A$ and $Eq^B$, are used to do the MAC-layer decoding for A and B separately. There is no PHY-layer bridging or MAC-layer bridging.
\item \emph{NCMA system (Multi-User)} \\
    In full NCMA, both MUD decoder (either RMUD or SIC) and PNC decoder are used at the PHY layer. All three equation systems $Eq^A$, $Eq^B$, and $Eq^{A \oplus B}$ are used for MAC-layer decoding. In particular, both PHY-layer bridging and MAC-layer bridging are performed in the decoding process.
\end{enumerate}



\begin{figure}
\centering
\includegraphics[width=0.5\textwidth]{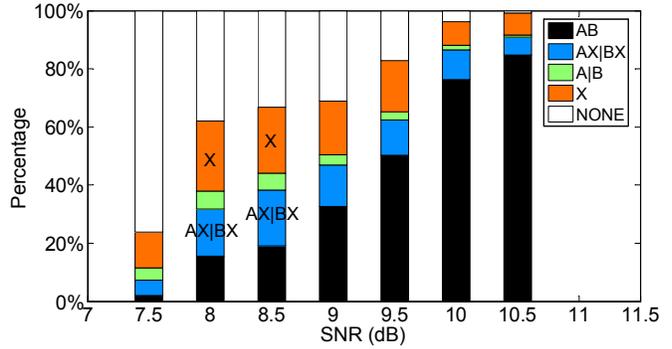}
\caption{NCMA PHY-layer packet decoding statistics versus SNR, with two nodes having the same SNR.}
\label{fig:NCMA-Exp-PHY-Distribution1}
\end{figure}

\subsubsection{PHY-Layer Packet Decoding Statistics}\label{sec:Exp21}

Our first experiment aims to gather PHY-layer decoding statistics. The results are summarized in Fig. \ref{fig:NCMA-Exp-PHY-Distribution1}. Recall that with the combined use of MUD and PNC decoders in NCMA, there are 8 possible outcomes (see Section \ref{sec:Overview2}). To ease presentation of the key results, we group some events in Section \ref{sec:Overview2} together by the following notations in Fig. \ref{fig:NCMA-Exp-PHY-Distribution1}:
\begin{itemize}\leftmargin=0in
\item \texttt{NONE} = (iv)(II) (no packet decoded).
\item \texttt{X} = (iv)(I) (only XOR packet decoded).
\item \texttt{A|B} = (ii)(II) + (iii)(II) (either only packet $A$ or only packet $B$ decoded).
\item \texttt{AX|BX} = (ii)(I) + (iii)(I) (XOR packet plus either packet $A$ or packet $B$ decoded).
\item \texttt{AB} = (i)(I) + (i)(II) (both packets $A$ and $B$ decoded; XOR packet may or may not be decoded).
\end{itemize}

We performed controlled experiments for different SNRs. The AP send 10,000 beacons to trigger simultaneous transmissions. The receive powers of nodes A and B at the AP are adjusted to be balanced, with less than 0.5dB difference between the two powers. We varied the ``common'' SNR from 7.5 dB to 10dB. The SNR calculation method is based on the scheme presented in \referred{HalperinSNR10}\cite{HalperinSNR10}. We performed the experiment 5 times for each SNR and average the results.

Recall from the discussion in Section \ref{sec:MacDec} that the complementary XOR packets are useful for PHY-layer bridging and the lone XOR packets are useful for MAC-layer bridging. As shown in Fig. \ref{fig:NCMA-Exp-PHY-Distribution1}, there are considerable complementary XOR packets (\texttt{AX|BX}) and lone XOR packets (\texttt{X}) across all SNRs. At 8.5dB, in particular, the events associated with complementary packets and lone packets both happen more than 20\% of the time. The contributions of these packets to the overall throughput of NCMA will become evident from the experimental results presented next.

\subsubsection{Overall NCMA Performance}\label{sec:Exp22}
Next, we evaluate the overall NCMA performance, with both PHY-layer decoding and MAC-layer decoding. For benchmarking, let us first derive a theoretical upper bound for the overall NCMA normalized throughput imposed by the PHY-layer decoding performance. In each time slot, depending on the event, we could obtain either one or two equations for use in our three equation systems at the MAC layer. For example, when both packets A and B are obtained by PHY-layer decoding, then the normalized throughput for that particular time slot is 2. The upper bound for the normalized NCMA throughput averaged over all time slots can be shown to be (note: lone XOR packet also counts as 1 packet):
\begin{align}
{\textnormal{Upper Bound}} = &2 \times \left(\Pr \{ \texttt{AB}\}  + \Pr \{ \texttt{AX|BX}\} \right) \nonumber\\
                     &+ 1 \times \left(\Pr \{ \texttt{A|B}\} + \Pr \{ \texttt{X}\} \right)
\label{equ:expupperbound}
\end{align}

This upper bound cannot exceed 2 since we allow at most two users to transmit together in our current prototype. It can be shown that even with MAC-layer bridging, this upper bound cannot be exceeded.


We next present experimental results showing that NCMA can achieve normalized throughput close to this upper bound. We employ trace-driven simulations for this set of experiments. Specifically, we first gather the PHY-layer event statistics and then use the traces to drive the events in our simulations. For NCMA, after each message is decoded, the associated node of that message begins sending another message, paired with the yet-to-be decoded message of the other node (see explanation in Section \ref{sec:MacDec25}).

\begin{figure}
\centering
\includegraphics[width=0.5\textwidth]{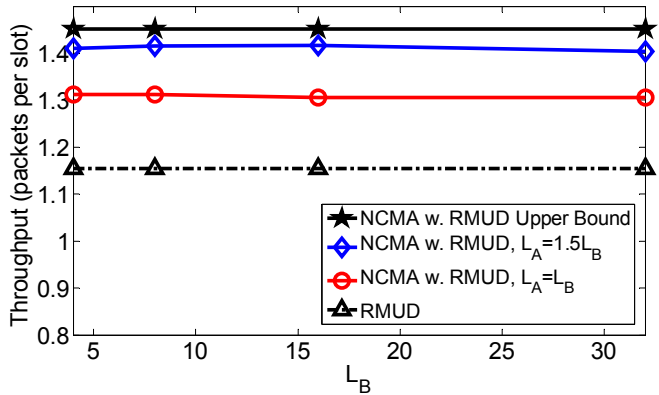}
\caption{Throughput comparison of different schemes with different RS code parameters, $L_B = 4, 8, 16, 32$, and fixed $SNR_A = SNR_B = 9$dB.}
\label{fig:NCMA-Exp-MAC-Throughput1}
\end{figure}



In Fig. \ref{fig:NCMA-Exp-MAC-Throughput1}, we show the normalized throughputs of various schemes at SNR = 9 dB. In the figure, $L_A$ ($L_B$) is the number of PHY-layer packets the MAC layer of node A (B) must have before each message $M^A$ ($M^B$) can be decoded. The normalized throughput is defined as $Th = {{N_A  + N_B } \over {N_{Beacon} }}$, where $N_A$ ($N_B$) and $N_{Beacon}$ are the total number of recovered MAC-layer packets from node A (B), and the total number of beacons, respectively. We can see from Fig. \ref{fig:NCMA-Exp-MAC-Throughput1} that NCMA with RMUD outperforms RMUD without the PNC decoder by 20\%,.



Note that making $L_A  \ne L_B $ improves performance. In particular, making $L_A  = 1.5L_B$ allows NCMA throughput to approach the upper bound. The reason that unequal $L$ is better is as follows. With respect to MAC-layer bridging, suppose that $M^B$ is decoded first. As explained in Section \ref{sec:MacDec2}, with the decoded $M^B$, MAC-layer bridging may allow us to obtain additional equations associated with $M^A$ through lone XOR packets. Each lone XOR packet gives us one more equation for $M^A$. When $L_A  = L_B$, the number of lone XOR packets may be ``more than enough'' (i.e., we end up having more than $L_A$ equations after bridging). The extra lone XOR packets are then wasted, because they do not contribute to the througput. By contrast, in the upper bound formula, each lone XOR packet contributes 1 unit to the throughput. Thus, each wasted lone packet pulls the throughput further away from the upper bound. Intuitively, we can see that it is better for $M^A$ not to be decoded at exactly the same time as $M^B$ , in which case all the lone XOR packets contribute to the throughput. Making $L_A  = 1.5L_B$ disaligns the decoding times of $M^A$ and $M^B$ and ensures this is more likely to be the case. Other disalignment strategies for the messages are also possible; to limit scope, we will not delve further into that direction here.


\begin{figure}
\centering
\includegraphics[width=0.5\textwidth]{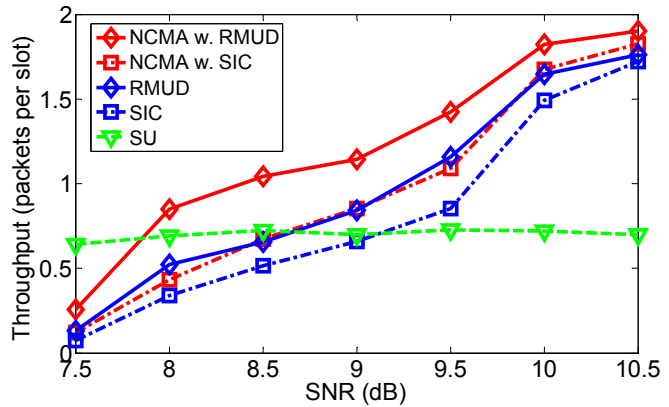}
\caption{Throughput comparison of different schemes for different SNRs ($SNR_A = SNR_B$) with $L_A  = 1.5 \times L_B = 24$.}
\label{fig:NCMA-Exp-MAC-Throughput-Diff-RS}
\end{figure}

We also see from Fig. \ref{fig:NCMA-Exp-MAC-Throughput1} that as long as $L_A = 1.5L_B$, the absolute value of $L_B$ is not important. For the next set of experiments, we fixed $L_B = 16$ and varied the SNR. Fig. \ref{fig:NCMA-Exp-MAC-Throughput-Diff-RS} shows that NCMA outperforms SU by 100\% when the average SNR$\ge$9.5dB. At SNR$\ge$9.5dB, NCMA has significantly better performance than RMUD and SIC without the PNC decoder. When SNR$>$10.5dB, all multiuser schemes (including MUD and NCMA) have good performance and they start to converge to throughput of 2.


Fig. \ref{fig:NCMA-Exp-MAC-Throughput-Diff-RS} also includes the results for NCMA with SIC. We see that its performance is not as good as that of NCMA with RMUD. This is understandable because Fig. \ref{fig:NCMA-Exp-MAC-Throughput-Diff-RS} concerns the balanced-power case, and SIC is known to have poor performance when powers from different users are balanced \referred{SICMobicom07}\cite{SICMobicom07}.

\subsubsection{Effects of Unbalanced-Power User Pairing}\label{sec:Exp23}

\begin{figure}[t]
\centering
\includegraphics[width=0.5\textwidth]{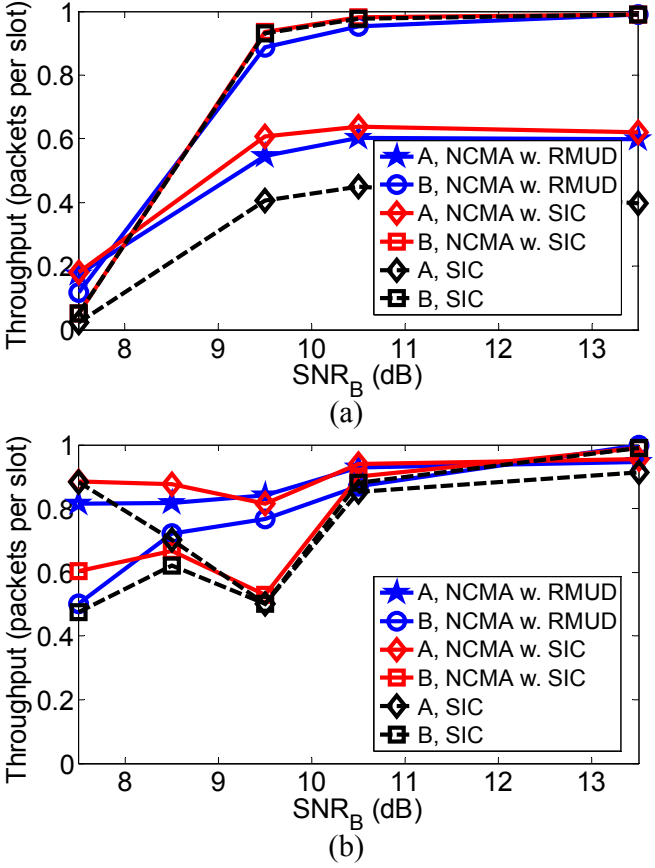}
\caption{Throughputs of node A and node B, with SNR of node A fixed at (a) $SNR_A = 7.5$dB and (b) $SNR_A = 9.5$dB, and SNR of node B varied. }
\label{fig:NCMA-Exp-MAC-Throughput-Diff-power}
\end{figure}

We next investigate what happens when the receive powers of users A and B are different. Our experiments indicate that NCMA is not only robust against power imbalance, but its performance can actually be better under unbalanced-power user pairing. Fig. \ref{fig:NCMA-Exp-MAC-Throughput-Diff-power} presents the results. The SNR of A was fixed to be 7.5dB and 9.5dB in Fig. \ref{fig:NCMA-Exp-MAC-Throughput-Diff-power}(a) and Fig. \ref{fig:NCMA-Exp-MAC-Throughput-Diff-power}(b), respectively. For each fixed power, we varied the SNR of B. Note that for a fixed $SNR_A$, as $SNR_B$ increases, not only the throughput of B improves, the throughput of A also improves. For example, when $SNR_A$ is fixed at 7.5dB, the throughput of A could be improved by 400\% as $SNR_B$ increases from 7.5dB to 8.5dB.

To better analyze the results, we adopt the following notations:
\begin{itemize}\leftmargin=0in
\item $(Th'_A, Th'_B) >> (Th_A, Th_B)$ means $Th'_A > Th_A$ and $Th'_B > Th_B$.
\item $(Th'_A, Th'_B) <> (Th_A, Th_B)$ means $Th'_A < Th_A$ and $Th'_B > Th_B$.
\item $(Th'_A, Th'_B) >< (Th_A, Th_B)$ means $Th'_A > Th_A$ and $Th'_B < Th_B$.
\item $Th^{NR} = (Th^{NR}_{A}, Th^{NR}_{B})$ stands for Throughputs of NCMA-RMUD.
\item $Th^{NS} = (Th^{NS}_{A}, Th^{NS}_{B})$ stands for Throughputs of NCMA-SIC.
\item $Th^{S} = (Th^{S}_{A}, Th^{S}_{B})$ stands for Throughputs of SIC.
\end{itemize}
Let us focus on Fig. \ref{fig:NCMA-Exp-MAC-Throughput-Diff-power}(b), where $SNR_A = 9.5$dB, for the analysis of the power imbalance benefits in NCMA. At $SNR_B = 7.5$dB, NCMA-SIC has the best performance, i.e., $Th^{NS} >> Th^{NR}$ and $Th^{NS} >> Th^{S}$. In particular, there is an appreciable gap between $Th^{NS}_{B}$ and $Th^{S}_{B}$, implying the use of PNC decoder that makes possible both PHY-layer bridging and MAC-layer bridging can improve the performance of the weak user significantly. At $SNR_B = 8.5$dB, both $Th^{NS} >> Th^{S}$ and $Th^{NR} >> Th^{S}$. Using SIC alone without the PNC decoder will result in inferior performance. Meanwhile, $Th^{NR} <> Th^{NS}$. This points to the possibility of using SIC and RMUD in combination at the PHY-layer decoding to improve performance. At $SNR_B = 9.5$dB, we have the balanced power case. Clearly, $Th^{NR} >> Th^{NS}$ and $Th^{NR} >> Th^{S}$, indicating that RMUD is a good complement to SIC so that the Achilles's heel of SIC, the balanced power case, can be overcome. At $SNR_B >= 10.5$dB, both NCMA-RMUD and NCMA-SIC have good performance. Overall, NCMA-RMUD has the ``smoothest performance'' (no large fluctuations in performance) when $SNR_B$ is varied.

\begin{figure*}
\centering
\includegraphics[width=1\textwidth]{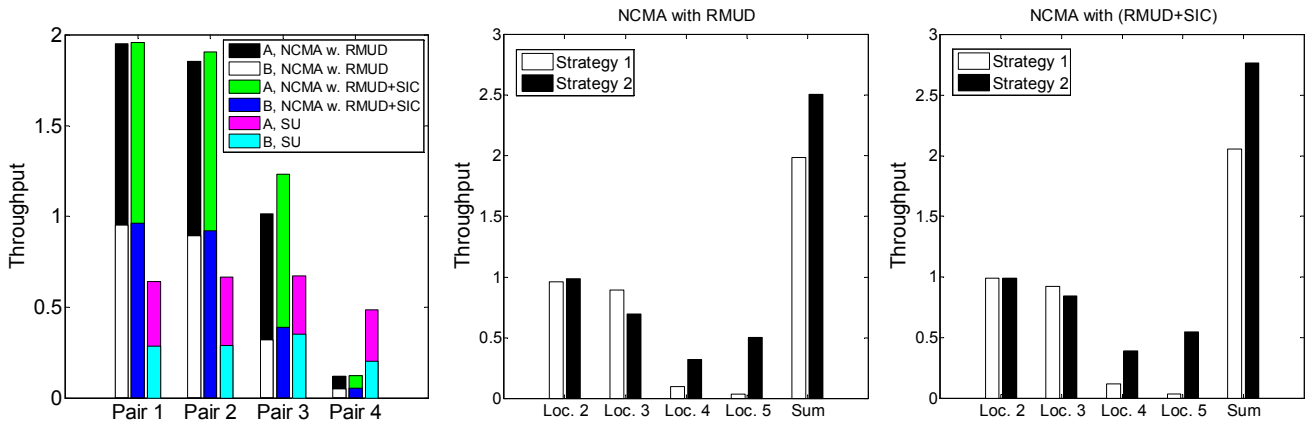}
\caption{Throughputs with AP placed at location 9 in Fig. \ref{fig:NCMA-Exp-Map}: (a) Throughputs of four user pairs, P1, P2, P3, and P4 in Table \ref{tab:NCMA-Pairing} under NCMA and SU; (b) Throughputs of two different user-pairing strategies under NCMA with RMUD; (c) Throughputs of two different user-pairing strategies under NCMA with (RMUD+SIC).}
\label{fig:NCMA-Exp-MAC-Throughput-Random1}
\end{figure*}

\begin{table}
\begin{normalsize}
\centering
\caption{User pairing in random topology}
\begin{tabular}{|c|l|l|} \hline
User Pair    &    User A            &   User B                 \\ \hline \hline
P1           &  Location 1 (20dB)   & Location 2 (12.3dB)      \\ \hline
P2           &  Location 2 (12.3dB) & Location 3 (9dB)      \\ \hline
P3           &  Location 3 (9dB)    & Location 4 (7dB)      \\ \hline
P4           &  Location 4 (7dB)    & Location 5 (7.4dB)      \\ \hline
P5           &  Location 2 (12.3dB) & Location 5 (7.4dB)      \\ \hline
\end{tabular}
\end{normalsize}
\label{tab:NCMA-Pairing}
\end{table}

\subsubsection{Which NCMA Variants to Use}\label{sec:Exp24-}

Our experimental results have validated the worthiness of introducing PNC decoding to our multiaccess system. Overall, NCMA-RMUD and NCMA-SIC, in which PNC decoding is used, have better performance than the corresponding RMUD and SIC, respectively.

A question is which variant of NCMA should be adopted. Fig. \ref{fig:NCMA-Exp-MAC-Throughput-Diff-power}(b) indicates that compared with NCMA-SIC, NCMA-RMUD has smoother performance transition from the unbalanced power setting to the balanced power setting. That is, it works well for both settings and is more robust if the system does not allow us to fine-tune the relative powers of user pairs. Furthermore, SIC requires a second round of PHY decoding after interference cancellation (which introduces processing delay), while RMUD performs all its decoding in one round. Thus, for a simple implementation that has low latency and good performance across all relative power settings, NCMA-RMUD is perhaps preferred. However, if complexity and latency are not an issue, then NCMA that uses both RMUD and SIC (rather than just one of them) should have the best throughput performance.

\subsubsection{Random Topology Experiments}\label{sec:Exp24}

We now present the experimental results of NCMA with RMUD, and NCMA with both RMUD and SIC, for a random topology. The random topology is constructed from the set-up in Fig. \ref{fig:NCMA-Exp-Map}, without deliberate power control, i.e., we placed the USRPs in different locations in an indoor environment, and each of them uses a fixed transmit power. We report the results of a particular set of experiments with the AP placed at location 9 in Fig. \ref{fig:NCMA-Exp-Map}. Users are placed at locations 1, 2, 3, 4, and 5. The associated SNRs for these locations are 20dB, 12.3dB, 9dB, 7dB, and 7.4dB, respectively. Five user pairs P1, P2, P3, P4, and P5 as shown in Table \ref{tab:NCMA-Pairing} are formed for experimentation purposes.

Fig. \ref{fig:NCMA-Exp-MAC-Throughput-Random1}(a) shows the throughputs of different user pairs P1, P2, P3, and P4, under NCMA and SU. We see that NCMA outperforms SU significantly, except for P4. The overall system throughput improvement of NCMA over SU is 100\%. In P1, P2, and P3, although the powers are unbalanced, at least one of the users has decent power. We also note that for the power-unbalanced P1, P2, and P3, NCMA with RMUD and SIC has slightly better performance than NCMA with RMUD. In P4, the powers of both users are low, a regime where multi-user decoders do not work well.

The above observation on P1, P2, P3, and P4 has the implication that it is better to pair a user with strong power with a user weak power. To better articulate this, consider a situation in which there are four users in the network at locations 2, 3, 4, and 5. We compare two strategies of user pairing. In \textsf{Strategy 1}, we form the user pairs P2 and P4 out of the four users; in \textsf{Strategy 2}, we form user pairs P3 and P5. As shown in Fig. \ref{fig:NCMA-Exp-MAC-Throughput-Random1}(b) and (c), \textsf{Strategy 2} has better overall performance. Note that in \textsf{Strategy 2}, we avoid pairing two weak users together as in P4 of \textsf{Strategy 1}. As a result, the throughputs of the two weak users at locations 4 and 5 are pulled up significantly by their stronger partners. Not only is \textsf{Strategy 2} more fair, its overall system throughput is also 20\% higher than that of \textsf{Strategy 1}.

\subsubsection{Further Improvements}\label{sec:Exp25}

We have presented the performance of our NCMA system based on the BPSK modulation, with SNRs ranging from 7dB to 10.5dB. The highest normalized throughput is bounded by 2 because at most two users are allowed to transmit concurrently. We believe that the throughput can be further increased by allowing more than two users to transmit together. We also note that when the SNRs are higher than 10dB, higher order modulations (e.g., 16-QAM) could be applied \referred{HalperinSNR10}\cite{HalperinSNR10} to better utilize the available power. For SNR below 7dB, our current decoders do not work well due to the approximation methods adopted. In this paper, we show that pairing a weak user with a strong user can substantially alleviate the problem. An alternative approach is to refine our approximation methods so that the decoders also work well when both users have low SNRs. The decoder extensions mentioned above await further work.

\section{Related Work} \label{sec:RelatedWork}

Network coding (NC) has been implemented and evaluated in wireless networks at the PHY layer \referred{PNC06, KattiANC07, FPNCPhycom12}\cite{PNC06, KattiANC07, FPNCPhycom12}, and the network layer \referred{KattiXORinAir06, KattiSymbolNC08, RoznerER07}\cite{KattiXORinAir06,KattiSymbolNC08, RoznerER07}. However, the previous studies of NC have generally been restricted to relay networks, where NC was originally shown to be helpful for packet exchange via relays. NCMA, on the other hand, targets the non-relay setting. NC has also found use in packet retransmission \referred{KattiSymbolNC08, RoznerER2007}\cite{KattiSymbolNC08, RoznerER07}. NCMA, by contrast, aims directly at packet transmission rather than retransmission.

Channel-coded PNC systems have been intensively studied in prior work (e.g., \cite{Nazer2011ReliablePNC,PNCSurveyPhycom12,ShengliJSAC09,APNCTWC12}). However, channel coding only appeared in the PHY layer in this prior work. By contrast, the overall NCMA system has MAC-layer channel coding as well as PHY-layer channel coding. In this paper, we assume the use of a convolutional code at the PHY layer and a simplified PNC decoder. The use of more sophisticated PNC decoders, such as those described in \cite{Nazer2011ReliablePNC,PNCSurveyPhycom12}, are also possible in NCMA.

Refs. \referred{CRMA, PNCAloha, ANCMAC}\cite{CRMA11, ANCMAC11} explored how to resolve collisions among packets of different users. Instead of discarding the collided packets, a set of linear equations are formed to exploit information contained in them. The decoding is based on PHY-layer equations only. In addition, the decoding methods are either pure MUD or pure PNC methods. By contrast, NCMA introduces correlations among PHY packets so that another layer of MAC decoding can be used to improve performance. Also, NCMA makes use of both MUD and PNC at the PHY layer in a complementary way.

In OFDMA, different users transmit their signals on different subcarriers \referred{JelloNSDI10, FICA10}\cite{JelloNSDI10, FICA10}. By contrast, in NCMA, multiple users use the same set of subcarriers when they transmit concurrently. This improves spectrum efficiency. In \referred{RepeatErran10}\cite{RepeatErran10}, there could be overlapped subcarriers between two users; however, at least some of the subcarriers  must be non-overlapping for successful packet recovery. Similarly, as a time-domain WLAN system, the set-up in \referred{Zigzag08}\cite{Zigzag08} requires some symbols to be non-overlapping in time to bootstrap the packet recovery algorithm.

Recently, interference cancellation techniques have been advanced and applied to the decoding of PHY-layer rateless codes \referred{Spinal2012, Strider2011, AutoMAC12}\cite{Spinal2012, Strider2011, AutoMAC12}. For these techniques, the processing is entirely on signal samples rather than on bits. While having good performance, the decoding procedure could incur considerable storage and computation costs. NCMA opts for reduced complexity for simple PHY-layer decoding with real-time performance. The correlations among different PHY packets are exploited in MAC-layer decoding, which deals with bits rather than samples. Our design is largely compatible with the processing flows of the current wireless standards (e.g., 802.11).

\section{Conclusions}\label{sec:Conclusions}
We have conceptualized and experimentally demonstrated a first WLAN system that jointly exploits PNC and MUD to decode concurrent transmissions by multiple users. Throughput gain of 100\% relative to the traditional single-user transmission system has been achieved at medium SNR (10dB). We believe our proposed concept has ample room for further advances. Interesting future investigations include:

\begin{itemize}\leftmargin=0in
\item \emph{Higher Throughput} -- In the current NCMA prototype, beyond SNR of 10 dB, we are already approaching the best possible normalized throughput of 2. This is because the system lets at most two users transmit concurrently. For higher normalized throughput, a possibility is to let more than two users transmit simultaneous. Also, our current system makes use of the BPSK modulation. When SNR is high, higher-order QAM modulations could be considered. The extension of our PNC and RMUD decoders based on the principle of reduced constellation for higher-order QAM remains to be investigated.
\item \emph{Better Performance in Low SNR Regime} -- Currently, the performance of our PHY decoders suffer at low SNR ($\le 7.5$ dB). This is because the approximations (see Section \ref{sec:PhyDec}) used in our decoders are tailored for the medium-high SNR regime. We believe the low-SNR performance of our MUD and PNC decoders can be improved through refinement of our approximations.
\item \emph{Combined Use with Other Advanced Techniques} -- Currently, distributed MIMO and full-duplex techniques are very active areas in wireless network research. Our current NCMA prototype does not make use of these techniques. NCMA is, however, complementary to these techniques. The combination of MIMO and full-duplex technologies with NCMA is a promising direction for further work.
\end{itemize}

\appendices
\section{Quantization of Soft Information} \label{sec:appendix1}
%

The quantized value of the soft information depends on the quantized values of the reduced constellation points. To set the quantized values of the constellation points, consider the noiseless case so that there are four possible distinct values for $y[k]$, corresponding to the four constellation points. The four constellation points need to be reduced to two constellation points before the standard VA can be used. Consider the PNC decoder. Suppose that the constellation point $(+1, +1)$ is transmitted so that $y[k] = h_A [k] + h_B [k]$. Then using the approximation method in Section \ref{sec:PhyDec},
\begin{align}
 \tilde{x}_{A \oplus B }[k]
&\approx \min \left(h_A^{} [k],h_B^{} [k]\right) \cdot \left( {y[k] - \max (h_A^{} [k],h_B^{} [k])} \right) \nonumber \\
& = \min \left(|h_A^{} [k]|^2 ,|h_B^{} [k]|^2 \right).
\label{equ:pncquant1}
\end{align}
Considering all the four constellation points under noiseless reception, it can be shown that there are only two reduced values for $\tilde{x}_{A \oplus B }[k]$, $ \pm \min (|h_A^{}[k]|^2, |h_B[k]|^2 )$. These two values are the two effective constellation points for bit $k$. Over all $k$, we can focus on the pair of reduced constellation points with the largest magnitude given by
\begin{align}
\pm |h_{\max } |^2  \buildrel \Delta \over = \pm \mathop {\max }\limits_k \min (|h_A^{} [k]|^2 ,|h_B^{} [k]|^2 ) . \label{equ:pncquant2}
\end{align}

For quantization, we first map $ \pm |h_{\max } |^2$ to two points on the real interval $[ - 0.5,0.5]$. Specifically, for a defined $\alpha  \in (0,0.5)$ (e.g., $\alpha  = 0.25$), we map $|h_{\max } |^2  \to \alpha $ and $ - |h_{\max } |^2  \to  - \alpha $.
Then, we normalize and quantize $\tilde{x}_{A \oplus B }[k]$ as
\begin{equation}
\tilde{x}_{A \oplus B, quantized }[k]   = \left( {{{\tilde{x}_{A \oplus B }[k] } \over {|h_{\max } |^2 }}\alpha  + 0.5} \right)\times255 .
\label{equ:pncquant3}
\end{equation}

For the MUD decoder that decodes A, using the same reasoning, we have
\begin{align}
\pm |h_{\max } |^2  \buildrel \Delta \over =  \pm \mathop {\max }\limits_k |h_A^{} [k]|^2 .  \label{equ:pncquant4}
\end{align}
\begin{align}
\tilde{x}_{A,quantized}[k]  = \left( {{{\tilde{x}_A [k] } \over {|h_{\max } |^2 }}\alpha  + 0.5}  \right)\times255  . \label{equ:pncquant5}
\end{align}
For packet $B$, we just replace $A$ by $B$ in the above formula.

We find that, empirically, optimal performance can be obtained at $\alpha $ between 0.2 and 0.25, and performance does not vary much within this range. Our experiments were conducted with $\alpha = 0.228$.

Quantizations of the RMUD and SIC decoders can be performed based on the same principle as above.

\ifCLASSOPTIONcompsoc
  \section*{Acknowledgments}
\else
  \section*{Acknowledgment}
\fi

This work is supported by AoE grant E-02/08 and the General Research Funds Project Number 414812, established under the University Grant Committee of the Hong Kong Special Administrative Region, China. This work is also supported by NSF of China (Project No. 61271277).



%


\bibliographystyle{IEEEtran}
\bibliography{ncma_sigproc}

\end{document}